\documentclass[apl,aip,twocolumn,reprint,floatfix,superscriptaddress]{revtex4}
\usepackage{graphicx}
\usepackage{subfigure}
\usepackage{amsmath}

\usepackage{natbib}

\newcommand{\fref}[1]{Fig.~\ref{#1}}
\newcommand{\fsref}[1]{Figs.~\ref{#1}}

\newcommand{\eref}[1]{Eq.~\ref{#1}}

\begin{document}

\title{Circular Photogalvanic Effect in Organometal Halide Perovskite CH$_3$NH$_3$PbI$_3$}
\author{Junwen Li$^{1,2}$, Paul M. Haney}
\affiliation{Center for Nanoscale Science and Technology, National Institute of Standards and Technology, Gaithersburg, MD 20899, USA \\
$^2$  Maryland NanoCenter, University of Maryland, College Park, MD 20742, USA }
\begin{abstract}
We study the circular photogalvanic effect in the organometal halide perovskite solar cell absorber CH$_3$NH$_3$PbI$_3$.  For crystal structures which lack inversion symmetry, the calculated photocurrent density is about $10^{-9}$ A/W, comparable to the previously studied quantum well and bulk Rashba systems.  Because of the dependence of the circular photogalvanic effect on inversion symmetry breaking, the degree of inversion asymmetry at different depths from the surface can be probed by tuning the photon energy and associated penetration depth.  We propose that measurements of this effect may clarify the presence or absence of inversion symmetry, which remains a controversial issue and has been argued to play an important role in the high conversion efficiency of this material.
\end{abstract}
\date{\today}
\maketitle

Organometal halide perovskites are emerging thin-film photovoltaic materials which can be fabricated with solution methods and can also exhibit remarkable power conversion efficiency. Both ABX$_3$ (A: CH$_3$NH$_3$, HC(NH$_2$)$_2$; B: Pb, Sn; X: Cl, Br, I) and their hybrids have exceptional photovoltaic performance and also show promise as LEDs, lasers and X-ray detectors.\cite{tan2014bright,xing2014low,yakunin2015detection}  The mostly studied CH$_3$NH$_3$PbI$_3$  perovskite exhibits three different structural phases from high to low temperatures with phase transitions  occurring at 327.4 K from cubic  to  tetragonal  and then at 162.2 K to orthorhombic.\cite{Kawamura_jpsj_2002}  In spite of the rapid progress in the power conversion efficiency, reaching more than 20~\% since its first application in 2009, a basic question about the crystal structure persists regarding the presence or absence of inversion symmetry. Stoumpos \textit{et al.} proposed that the crystal belongs to the noncentrosymmetric $I4cm$ space group with a ferroelectric distortion and octahedra rotation.\cite{Stoumpos_ic_2013} The ferroelectricity was used to explain the observed hysteresis in $J$-$V$ curve. First-principles calculations suggest that a ferroelectric distortion could be stable.\cite{Frost_nl_2014,zheng2015rashba,amat2014cation} In addition, Rashba spin-splitting of the electronic band structure due to inversion symmetry breaking was observed in angle resolved photoemission measurements.\cite{niesner2016giant} However, there is also a significant body of work which indicates that the material is centrosymmetric. The hysteretic behavior has been ascribed to charge trapping at the surface and ionic migration under applied bias.\cite{beilsten2015non, tress2015understanding,meloni2016ionic,ADMA:ADMA201503832,chen2016origin}  The crystal structure has been assigned to the centrosymmetric $I4/mcm$ space group\cite{poglitsch1987dynamic,Kawamura_jpsj_2002,weller2015complete} and centrosymmetry was assumed to explain the observed spin dynamics.\cite{Giovanni_nl_2015}   Light-induced structural changes could further complicate the assignment of the crystal space group.\cite{Liu_jpcl_2016, coll2015polarization,wu2015composition,gottesman2015photoinduced}

\begin{figure}[!htp]
 \includegraphics[angle=0,scale=0.58]{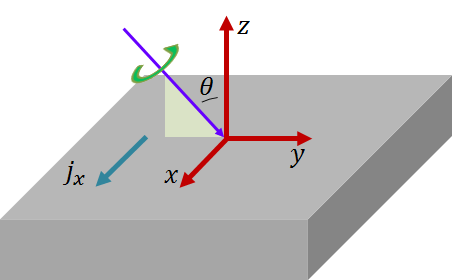}
  \caption{ Schematic of the circular photogalvanic measurement. The incident light with right circular polarization is propagating within the $\hat{y}$-$\hat{z}$ plane (indicated by the light green triangle) with incidence angle $\theta$. Inversion symmetry breaking is assumed along $\hat{z}$ direction. Photocurrent is flowing perpendicular to the plane of incidence. In our model system, the photocurrent is along the $\hat{x}$ direction.}
  \label{fig:schematic}
\end{figure}

In this paper, we propose that the circular photogalvanic effect  -- which is only allowed in gyrotropic media~-- could shed  light on the symmetries of the  space group of tetragonal CH$_3$NH$_3$PbI$_3$. \fref{fig:schematic} shows the cartoon of the circular photogalvanic effect measurement setup in which circularly polarized light induces a photocurrent. The photocurrent measurement yields information about the absence or presence of inversion symmetry breaking. Moreover, the surface and bulk are expected to exhibit different magnitudes of inversion symmetry breaking. This difference could be detected by varying the photon energy and associated absorption depth. As we discuss in more detail, a larger photon energy probes the material response in closer proximity to the sample surface, where inversion symmetry breaking is generally expected to be larger. In our previous study, we have shown the inversion symmetry breaking leads to an interesting and potentially useful optical spin response of the perovskites, which can be used to study the role of grain boundaries.\cite{Li_prb_2016}

\begin{figure}[!htp]
\includegraphics[angle=0,scale=0.4]{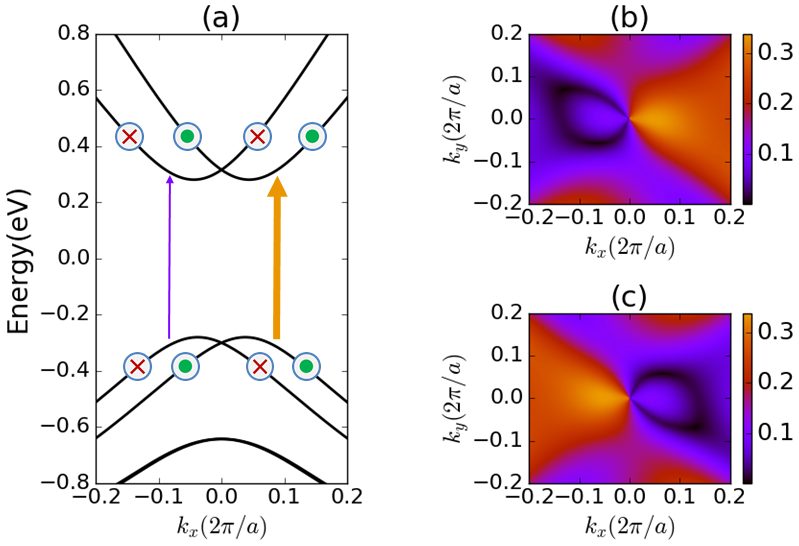}
  \caption{ (a) depicts the Rashba-type band structure of CH$_3$NH$_3$PbI$_3$ along $k_x$ ($k_y = 0$, $k_z = 0$). Circles with cross~(red) and dot~(green) indicate the spin orientations pointing into and out of the paper, respectively. The arrows with different thickness represent the asymmetric transition rates at ${\bf k}$ and $-{\bf k}$ for right circularly polarized light. (b) and (c) show matrix elements $\frac{1}{2} | \langle \psi_\textrm{c}\left ( {\bf k}\right ) | \hat{p}_x \pm i\hat{p}_z| \psi_\textrm{v} \left( {\bf k} \right ) \rangle |^2$ for transitions between highest valence band and lowest conduction band. $+$ and $-$ correspond to right and left circularly polarized light, respectively.  All molecules are arranged initially along $\hat{z}$ direction to obtain a larger inversion symmetry breaking. }
  \label{fig:transition}
\end{figure}

Herein, we report a first-principles density functional theory study on the circular photogalvanic effect focusing on noncentrosymmetric tetragonal CH$_3$NH$_3$PbI$_3$.  The absence of inversion symmetry, together with the strong spin-orbit coupling in the heavy element Pb, leads to the Rashba-type spin splitting as depicted in \fref{fig:transition} (a). In this case, the spin has a preferred orientation perpendicular to both the momentum and the inversion symmetry breaking direction.\cite{even2014dft} Because of  spin-momentum locking and angular momentum selection rules, optical transitions between conduction and valence energy states respond differently to light with different circular polarization direction. The asymmetric distribution of excited charge carriers leads to a photocurrent in the absence of an external bias. The symmetry properties of the photocurrent response are determined by the symmetry of the system Hamiltonian. For a Rashba model, the photocurrent direction is normal to the plane formed by the light angular momentum and the bulk symmetry breaking direction.\cite{ganichev2003spin} A characteristic feature of circular photogalvanic effect is that incident light with right and left circular polarization will lead to photocurrents flowing along opposite directions.

\fsref{fig:transition} (b) and (c)  show the optical transition amplitudes versus wave vector for right and left circularly polarized light. The transitions exhibit strong ${\bf k}$ dependence and energy states with $k_x > 0$~($< 0$) make more contribution to the photocurrent for right~(left) circularly polarized light. Moreover, the transition at ${\bf k}$ for right circular polarization is equal to that at $-{\bf k}$ for left circular polarization, indicating the characteristic dependence of photocurrent on circular polarization. This effect has been observed in bulk Te, GaAs and InAs quantum wells as well as bulk GaAs subjected to an external magnetic field.\cite{asnin1978observation,Ganichev_prl_2001,ivchenko2012superlattices,ganichev2003spin} It was also used to demonstrate the Rashba spin splitting in GaN-based heterostructures and to detect the lattice polarity of InN.\cite{weber2005demonstration,zhang2009lattice} Recently this effect was observed in bulk Rashba system BiTeBr\cite{ogawa2014photocontrol} and theoretical studies have been reported on an ultrathin film of topological insulators\cite{Wu_physicae_2012} and on graphene deposited on heavy-element substrates.\cite{inglot2015enhanced}

Because the spin-orbit coupling effect is dominated by the heavy element Pb, the molecular orientation and the ensuing distortion of PbI$_6$ octahedron plays an important role in determining the symmetry breaking.\cite{quarti2014interplay} We first study the model system with all molecules initially arranged along $\hat{z}$ direction to obtain a larger inversion symmetry breaking. The density functional theory calculations are carried out using local density approximation in the form of norm-conserving pseudopotentials as implemented in Quantum-ESPRESSO\cite{Paolo_jpcm_2009} with an energy cutoff of 40 $E_h$~( $1~E_h = 27.21$~eV ) for the plane wave basis expansion. A $4 \times 4 \times 3$ grid for the Brillouin zone sampling was employed during the structural relaxation and all atoms in the unit cell were allowed to move until the force on each  is less than 0.5 eV/nm. Troullier-Martins norm-conserving pseudopotentials\cite{Troullier_prb_1993, Engel_prb_2001} for all of the elements were generated with the APE, the Atomic Pseudo-potentials Engine.\cite{Oliveira_cpc_2008}  The lattice constants are calculated to be $a$ = 0.875~nm and $c$ = 1.203~nm, in good agreement with the experimental measurement (\hbox{$a$ = 0.880~nm} and \hbox{$c$ = 1.269~nm})\cite{Kawamura_jpsj_2002}. To remedy the underestimation of the energy gap in local density approximation, the calculated curve has been rigidly shifted to match an experimental energy gap of 1.5 eV. The tetragonal lattice structure we consider is noncentrosymmetric and exhibits ferroelectricity. The polarization is calculated to be 10.7~$\mu$C$/$cm$^2$ using Berry phase approach.

By considering the response of the system to a monochromatic electric field of frequency $\omega$
\begin{equation}
\mathbf{E}(t) = \mathbf{E}(\omega) e^{-i\omega t} + \mathbf{E}^*(\omega) e^{i\omega t} \rm,
\end{equation}  we solve semiconductor optical Bloch equations perturbatively to first order in the field intensity\cite{Schafer_springer_2002} and derive the photocurrent generation rate $\mathbf{\dot{J}}$  given by

\begin{equation}
\dot{J}^{i} = \chi^{ilm}(\omega) E^{l*}(\omega) E^{m}(\omega) \rm, \label{eq:qdot}
\end{equation}
where
\begin{equation}
\begin{aligned}
\chi^{ilm}(\omega) & = \frac{2\pi e^2}{\hbar^2 \omega^2} \frac{1}{V} \sum_{\mathbf{k}} \sum_{\textrm{c,v}} ( J^{i}_\textrm{c} - J^{i}_\textrm{v} )   \\
& \times v^{l*}_{\textrm{cv}} (\mathbf{k}) v^{m}_{\textrm{cv}}(\mathbf{k})
 \delta[ \omega_{\textrm{cv}}(\mathbf{k}) - \omega ] \label{eq:mu1} \rm.
\end{aligned}
\end{equation}

In \eref{eq:mu1}, $v^{m}_{\textrm{cv}}\left({\bf k}\right)=\langle \psi_\textrm{c}\left({\bf k}\right) | \hat{v}^m | \psi_\textrm{v}\left({\bf k}\right) \rangle$ is the velocity operator matrix element  between conduction and valence band states, $J^{i}_{\textrm{c}(\textrm{v})}=\langle \psi_{\textrm{c}(\textrm{v})}\left({\bf k}\right) | e \hat{v}^{i} | \psi_{\textrm{c}(\textrm{v})}\left({\bf k}\right) \rangle$, and $\omega_{\textrm{cv}}({\bf k})= \left(E_{\textrm{c}\bf k} - E_{\textrm{v\bf k}} \right)/\hbar$.  $\psi_{\textrm{c}(\textrm{v})}\left({\bf k}\right)$ represents the wave function of conduction~(valence) band with energy $E_{\textrm{c(v)}{\bf k}}$. The superscripts $i,l$ and $m$ indicate Cartesian components and summation over repeated indices is implied.  We first calculate the energies and momentum matrix elements on a coarse grid in momentum space and  employ Wannier interpolation technique\cite{Giustino_prb_2007,Yates_prb_2007,Marzari_rmp_2012}  to evaluate the photocurrent response (\eref{eq:mu1}) on a fine grid of $100 \times 100 \times 75$ $k$-points. This method has been applied in the study of optically injected spin current.\cite{Li_prb_2016}

\begin{figure}[!htp]
\subfigure[]{\includegraphics[angle=0,scale=0.156]{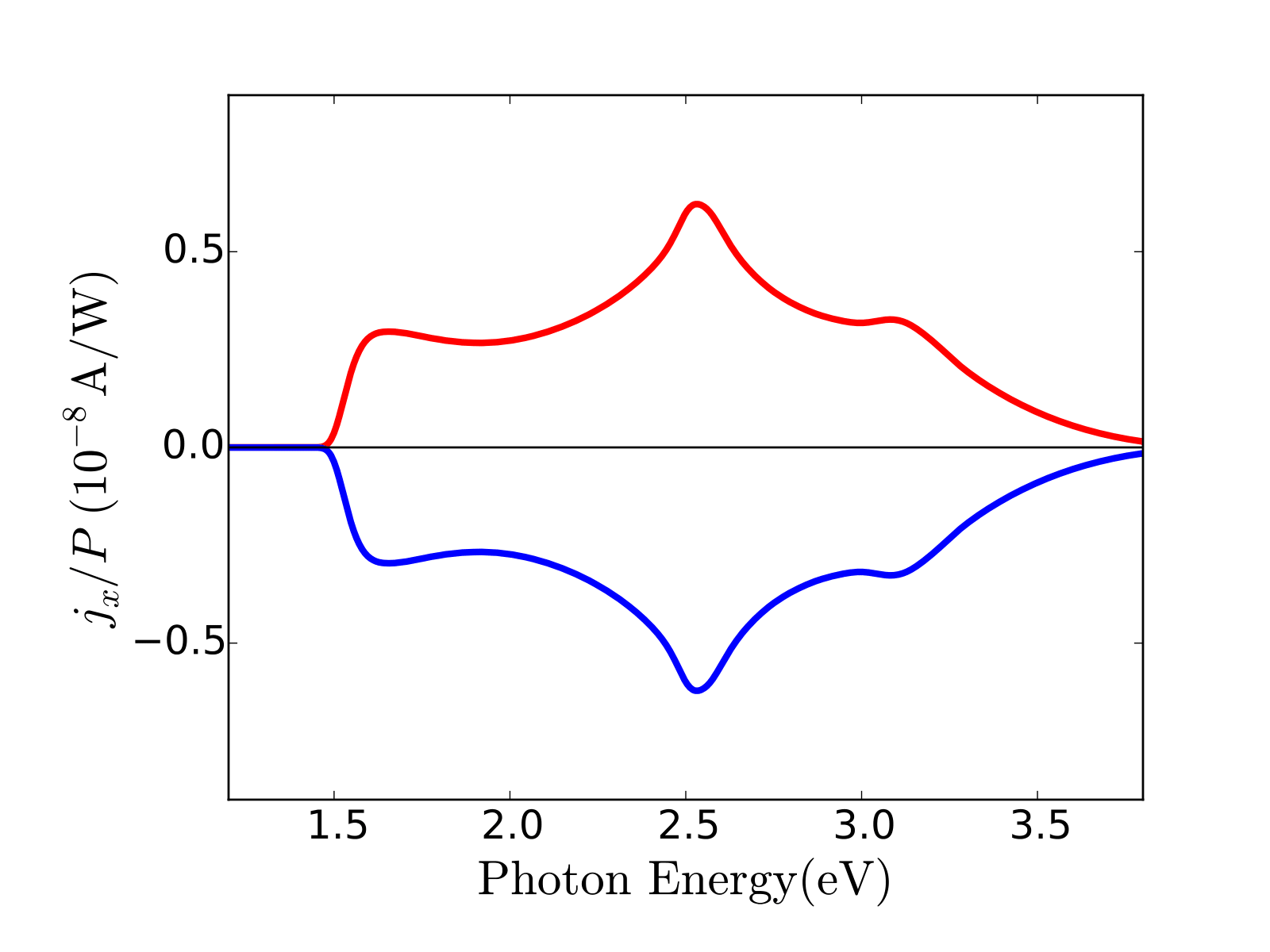}}
\subfigure[]{\includegraphics[angle=0,scale=0.156]{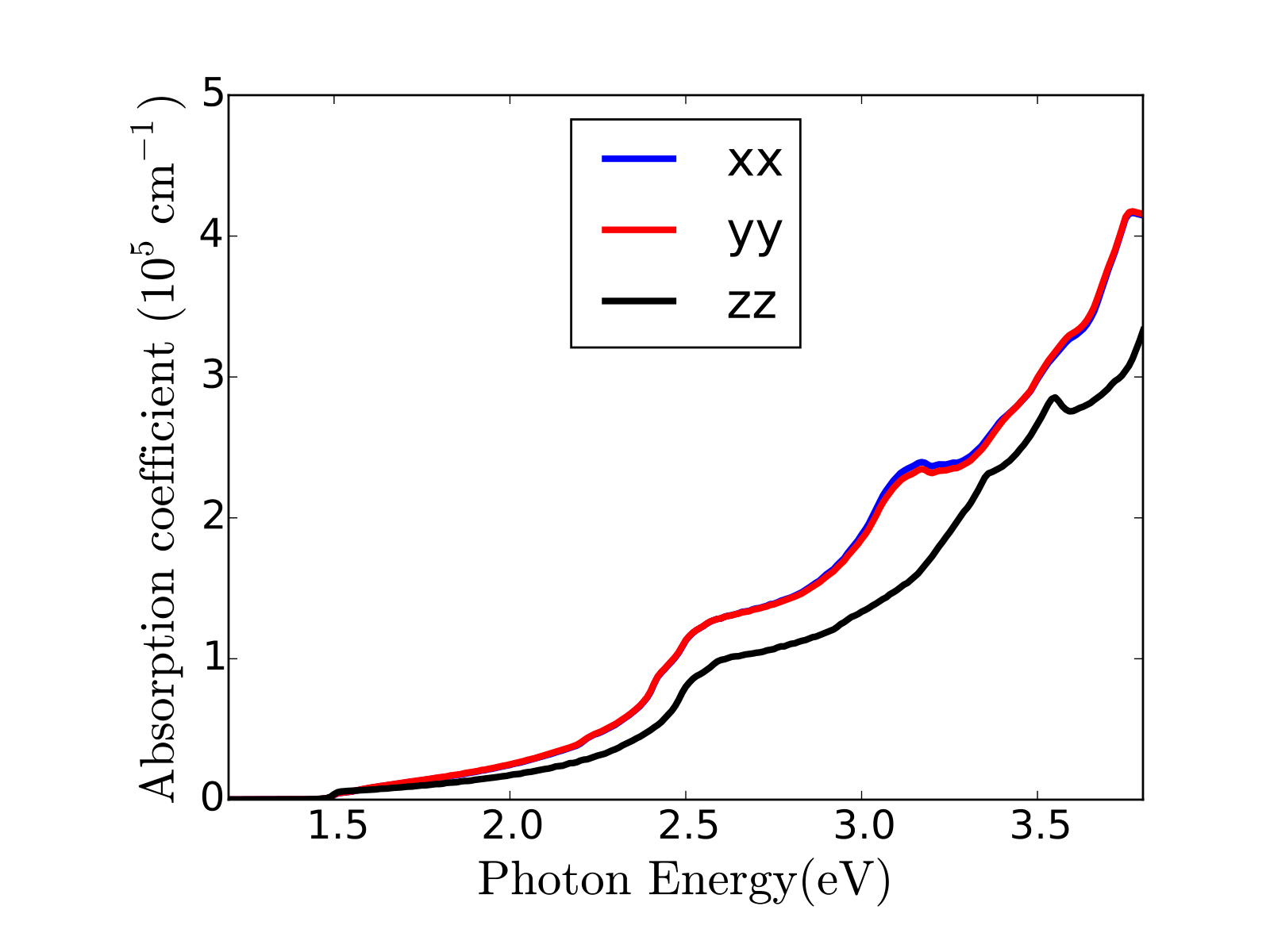}}
  \caption{ (a) Photocurrent density $j_x$ normalized to the radiation power $P$ as a function of the incident photon energy $\hbar\omega$ with respect to the energy gap for an incidence angle $\theta =90^o$. Positive (Negative) photocurrent direction corresponds to right (left) circularly polarized light. (b) Absorption coefficients for different polarizations as a function of incident photon energy. The close similarity between absorption coefficients for $xx$ and $yy$ polarizations corresponds to the symmetry of the system under study.}
  \label{fig:cpge_response}
\end{figure}

Since the system under study assumes inversion symmetry breaking along $\hat{z}$ direction, the spin orientation is confined in the $\hat{x}$--$\hat{y}$ plane as expected from the Rashba model. Neglecting reflection from the surface, we find  the largest response corresponds to the circularly polarized light incident along the $\hat{y}$ direction (incidence angle $\theta = 90^o$, see \fref{fig:schematic}) with electric field $\mathbf{E}(\omega) = \frac{1}{\sqrt{2}} E^0 (\hat{x} \pm i \hat{z})$ where $+$ and $-$ correspond to right and left circular polarization, respectively.  In the momentum relaxation time approximation under direct optical transitions we have
$ j_x = \dot{J}^x \tau \rm. $\cite{ganichev2003spin}
We assume the momentum relaxation time $\tau = 1$~fs for both electrons and holes. \fref{fig:cpge_response} (a) depicts the calculated ratio of photocurrent density $j_x$ to radiation power $P$ as a function of photon energy $\hbar\omega$ and clearly shows the switching of photocurrent direction upon the circular polarization reversal, a characteristic feature of the circular photogalvanic effect.  The peak value is about  $0.6 \times 10^{-8} \mbox{A/W}$, which is comparable with the previously reported photocurrents in $n$-InAs and $p$-GaAs quantum wells\cite{ganichev2003spin} and bulk Rashba semiconductor BiTeBr\cite{ogawa2014photocontrol}.

Owing to the dependence of circular photogalvanic effect on the inversion symmetry breaking, this effect could be used to detect the crystal structure and some of the associated mechanisms that have been proposed to explain the high power conversion efficiency of perovskite materials. In the organometal halide perovskites, a long carrier lifetime of about 300 ns has been observed.\cite{Stranks_science_2013,Xing_science_2013,wehrenfennig2014high} This has been attributed to the reduced nonradiative recombination by unusual defect physics\cite{yin2014unusual} and photon cycling\cite{Luis_science_2016}. In addition, the Rashba effect, for which inversion symmetry breaking is required, could have strong effect on the carrier lifetime. When conduction and valence bands exhibit different spin helicity, the associated spin mismatch significantly reduces the radiative recombination rate.\cite{zheng2015rashba} The indirect band gap caused by the different momentum offsets in valence and conduction bands may also reduce radiative recombination.\cite{zheng2015rashba, motta2015revealing, azarhoosh2016relativistic} Note that  circular photogalvanic effect is sensitive to the macroscopic inversion symmetry breaking, whose absence cannot rule out the possibility of reducing recombination by Rashba splitting. The local symmetry breaking or dynamic  Rashba splitting may contribute to long carrier lifetimes.\cite{zheng2015rashba,Etienne_jpcl_2016}

\def\kvec{ {\bf k} }
\def\evec{ {\bf e} }
\def\pvec{ {\bf \hat{p}} }
\def\Mvec{ {\bf M} }
\def\rvec{ {\bf r} }
\def\Gvec{ {\bf G} }
\def\eck{ E_{\textrm{c}\kvec} }
\def\evk{ E_{\textrm{v}\kvec} }

To further investigate how we can probe the inversion symmetry breaking and ensuing photocurrent response at different depths from the surface, we calculated the absorption coefficient as a function of the incident photon energy $\hbar\omega$ as depicted in \fref{fig:cpge_response} (b). The absorption coefficient $\alpha(\omega)$ is related to the complex dielectric function via
\begin{align}
\alpha(\omega) &= \frac{2\omega}{c} \sqrt{ \frac{ \sqrt{\varepsilon^2_1(\omega) + \varepsilon^2_2(\omega)} - \varepsilon_1(\omega) }{2} } \rm{,}
\end{align}
where $\varepsilon_1(\omega)$ and $\varepsilon_2(\omega)$ are the real and imaginary parts of the complex dielectric function, respectively. $\varepsilon_2(\omega)$ is given by
\begin{align}
\varepsilon_{2}^{\alpha\beta} (\omega)
 & =  \frac { e^2 \hbar^2 } { \pi m^2 }
\sum_{\textrm{c,v}} \int{d\kvec }
     \frac { p_{\textrm{cv}}^\alpha (\kvec) p_{\textrm{cv}}^\beta ( \kvec )^*}
           { \left( \eck - \evk \right)^2 }
           \delta ( \eck - \evk - \hbar\omega )
\end{align}
where the integral is over the first Brillouin Zone and the sum runs over every couple of valence and conduction bands. $p_{\textrm{cv}}^{\alpha(\beta)}(\kvec)$ denotes the dipole matrix element $\left< \psi_\textrm{c}(\kvec)|\hat{p}_{\alpha(\beta)}|\psi_\textrm{v}(\kvec) \right>$ representing
 the probability amplitude of the transition from state $\psi_\textrm{v}(\kvec)$ to state $\psi_\textrm{c}(\kvec)$, where  $\hat{p}$ is the momentum operator. The real  part of the dielectric function $\varepsilon_1(\omega)$ is calculated from $\varepsilon_2(\omega)$ through the Kramers-Kronig relation.

The calculated absorption coefficient is comparable to the experimental value.\cite{Wolf_jpcl_2014}  Near the band edge it is about $ 10^4 \, \mbox{cm}^{-1}$ and increases monotonically with photon energy. This energy dependence can be attributed to the density of states, with  Pb-$s$ orbitals in the valence band and Pb-$p$ orbitals in the conduction band. Because the dominant optical transition is between these intra-atomic orbitals of Pb atoms, more transitions contribute to the absorption when the photon energy increases.

The penetration depth, characterizing the depth at which the intensity of the radiation decays to $1/e$ of its surface value, is the inverse of the absorption coefficient.  \fref{fig:cpge_response} (b) indicates that the absorption depth goes from 1000 nm for photon energies near the band gap, to 50 nm at 3 eV. This enables a systematic study of the inversion asymmetry at different depths away from the surface. The surface and bulk dominate the photocurrent for large and small photon energies, respectively.  The nearly zero photocurrent for large photon energy would be indicative of the absence of bulk inversion symmetry breaking. If we assume inversion symmetry is only locally broken at the sample surface, then the electronic structure acquires Rashba-like features over a distance $L$ from the surface.\cite{haney2013current}  For metals, $L$ is only a few atomic layers, but in semiconductors, $L$ can be as large as the depletion width.  For an absorption coefficient $\alpha(\omega)$, the circular photogalvanic effect signal contains a factor of $(1- e^{-\alpha(\omega)L})$.  In this case a signal is only measured if the absorption occurs within a length scale of $L$.

In order to observe photocurrent response when the inversion symmetry breaking is due to the sample surface, oblique  incidence ($\theta \neq 0$) is required. As shown in  \fref{fig:schematic}, the inversion symmetry breaking is assumed to be normal to the surface. Considering the strong potential from structural asymmetry, we for simplicity assume that the $\hat{x}$--$\hat{y}$ plane is isotropic and inversion symmetric. When the light propogates with $\theta = 0$, the in-plane symmetry leads to the cancellation for the photocurrent contributions at ${\bf k}$ and $-{\bf k}$ and therefore, zero net photocurrent. The photocurrent first increases with incidence angle and then decreases due to enhanced portion of reflected light. When photon energy is smaller than 3 eV, the calculated refractive index varies between 2.1 and 2.59, from which we estimate the photocurrent response reaches its maximum at  $\theta \approx 45^o$.\cite{ganichev2003spin} However, distortion transverse to the surface normal may be present. McLeod \textit{et al.} found that perovskite thin films made by the one-step deposition method exhibit angle-dependent features in X-ray absorption spectroscopy measurements, indicating long-range alignment of the dipolar CH$_3$NH$_3$ molecules parallel to the surface.\cite{Mcleod_jpcl_2014}  In this case, the photocurrent response deviates from our prediction based on the \hbox{in-plane} isotropy and  inversion symmetry breaking within the plane can be detected with a nonzero photocurrent response for $\theta = 0$.

Inversion symmetry breaking could play an important role in the properties of photovoltaic materials. Its presence in the bulk region would be indicative of the role of Rashba splitting in reducing the radiative recombination.\cite{zheng2015rashba} Even inversion symmetry breaking only near the surface provides an effective route of investigating the charge current distribution near the grain boundaries in polycrystalline perovskite absorbers.  Because of the Rashba splitting, a nonequilibrium spin density is developed and is aligned in a direction perpendicular to the symmetry breaking direction and particle velocity.~\cite{Li_prb_2016} Measuring the spin density could therefore indicate the velocity direction of carriers, which could in turn help  elucidate the role of grain boundaries and other defects on charge transport.  Surface measurements, such as magneto-optical Kerr microscopy, probe the response in the inversion asymmetric environment near the surface, and could measure these effects even if the bulk is inversion symmetric. Niesner \textit{et al.} has performed the angle-resolved photoelectron spectroscopy on the surface valence bands of CH$_3$NH$_3$PbBr$_3$ and found that the Rashba strength is among the largest values reported so far.\cite{niesner2016giant}

In summary, we report on first-principles density functional study on the optical generation of photocurrent with circularly polarized light in the organometal halide perovskite CH$_3$NH$_3$PbI$_3$. We found that the photocurrent response is comparable with that of previously studied quantum well and bulk Rashba systems. Because of its dependence on the inversion symmetry breaking, the circular photogalvanic effect could find application to examine whether the sample under study is centrosymmetric or noncentrosymmetric which is critical for the understanding of the long carrier lifetime and therefore, the remarkable power conversion efficiency.

J. L. acknowledges support under the Cooperative Research Agreement between the University of Maryland and the National Institute of Standards and Technology Center for Nanoscale Science and Technology, Award 70NANB10H193, through the University of Maryland.

\bibliography{ref}

\end{document}